\documentclass[twoside]{article}
\usepackage{fleqn,espcrc2}
\usepackage[dvips]{graphicx}

\newcommand{\AmS}{{\protect\the\textfont2
  A\kern-.1667em\lower.5ex\hbox{M}\kern-.125emS}}
\title{Proton-Proton Total Cross Sections from the Window\\ 
of Cosmic Ray Experiments}

\author{A.A. Arkhipov\address{Theoretical Physics Division, 
        Institute for High Energy Physics, \\ 
        142281 Protvino, Moscow Region, Russia}} 

\begin{document}
\def\be{\begin{equation}}					 
\def\ee{\end{equation}}
\def\ber{\begin{eqnarray}}
\def\eer{\end{eqnarray}}	 		

\begin{abstract}
The importance of cosmic-ray experimental measurements of
proton-proton total cross sections to understand the underlying
fundamental dynamics is discussed. It is shown that early discovered
global structure of proton-proton total cross section \cite{5,17} is
completely compatible with the values obtained from cosmic-ray
experiments. 

\vspace{0.3pc}
\end{abstract}

\maketitle

PACS numbers: 11.80.-m, 13.85.-t, 21.30.+y

\vspace{0.3pc}
Keywords: total cross-sections, global structure, cosmic air showers,
numeric calculations, fit to the data,  interpretation of
experiments.
 
\section*{Introduction}
It is a well known fact that at energies above $\sqrt{s}\sim 20\,GeV$
all hadronic total cross sections rise with the growth of energy.
In 1970 the experiments at the Serpukhov accelerator revealed that
the $K^+p$ total cross section increased with energy \cite{1}.
Increase of the $pp$ total cross section has been discovered at the
CERN ISR \cite{2} and then the effect of rising $p\bar p$ total cross
sections was confirmed at the Fermilab accelerator \cite{3} and CERN
$Sp\bar p S$ \cite{4}.

Although nowadays we have in the framework of local quantum field
theory a gauge model of strong interactions formulated in terms of
the known QCD Lagrangian its relations to the so called ``soft"
(interactions at large distances) hadronic physics are far from
desired. In spite of 30 years after the formulation of QCD, we cannot
still obtain from the QCD Lagrangian the answer to the question why
and how all hadronic total cross sections grow with energy. We cannot
predict total cross sections in an absolute way starting from the
fundamental QCD Lagrangian as well mainly because it is not a
perturbative problem. 

The behaviour of hadronic total cross sections at high energies is a
wide and much discussed topic in high-ehergy physics community; see
e.g. the proceedings of famous Blois Workshops. At present time there
are a lot of different models which provide different energy
dependencies of hadronic total cross sections at high energies.

All different phenomenological models can conditionally be separated
into two groups in according to two forms of strong interaction
dynamics used: t-channel form and s-channel one \cite{5}. The first
group contains the Regge-type models with power-like,
$s^{\alpha_P(0)-1}$, behaviour of hadronic total cross sections. Here
$\alpha_P(0)$ is an intercept of the supercritical Pomerom
trajectory: $\alpha_P(0)-1=\Delta<<1$, $\Delta>0$ is responsible for
the growth of hadronic cross sections with energy; see recent paper
\cite{6} and references therein. There are a lot of people who works
with such a type of Regge-pole models.

However some part of scientific community works in the field related
to $s$-channel form of strong interaction dynamics and elaborates the
impact picture or geometrical models \cite{7,8,9,10,11,12,13}, which
exhibit an $ln^{2}s$ high-energy dependence and therefore it
asymptotically appears a saturation of the Froissart bound \cite{14}
in these models.

In our opinion the second group of the models is more preferable than
the first one from many points of view (see discussion in \cite{5}).
Moreover, careful analysis of the experimental data on hadronic total
cross sections and comparative study of two above mentioned
characteristic asymptotic parameterizations have shown that
statistically a ``Froissart-like" type parameterization for hadronic
total cross sections is strongly favoured
\cite{15,16}.\footnote{Recent investigations confirm this statement:
see e.g. COMPETE Collaboration e-print hep-ph/0107219, even though we
don't agree on many points of view suggested there.}

On the other hand if we suppose that unitarity saturation of
fundamental forces takes place at super-high energies then the energy
dependence of hadronic total cross sections can be derived and
investigated independently of phenomenological models but using only
general principles of relativistic quantum theory, such as
analyticity and unitarity, together with dynamic apparatus of
single-time formalism in QFT \cite{5,17}.

The experimental information on the behaviour of hadronic total cross
sections at ultrahigh energies can be obtained from cosmic ray
experiments. In this respect, analysis of extensive air showers
observations provides a unique source of such information. In fact,
the ultrahigh energy hadronic interactions occur when a primary
cosmic ray proton collides on air nucleus and as a result the
extensive air showers are produced by hadronic cascade in the
atmosphere. The primary cosmic ray protons with energy of
$10^{18}\,eV$ have been observed in the Utah ``Fly's Eye" detector.
This energy significantly exceeds the energy available at now working
accelerators and LHC in the near future as well. That is why, the
cosmic-ray data on hadronic total cross sections are most important.

\section{Cosmic-ray experiments and phenome\-no\-logy}

Recently we have two sets of data on proton-proton total cross
sections extracted from cosmic ray air showers observations
\cite{18,19}, including one point at $\sqrt{s}=30\,TeV$ from Fly's
Eye Collaboration experiment \cite{18} and six points up to
$\sqrt{s}=24\,TeV$ from AGASA Collaboration experiment \cite{19}.

It is well known fact that extracting proton-proton total cross
sections from cosmic ray extensive air showers observations is not so
straightforward. The physical description of extensive air showers
created by hadronic cascade in the atmosphere depends significantly
on the fundamental dynamics for hadron-nucleus and nucleus-nucleus
interactions at ultrahigh energies. Moreover, a procedure of
extracting any information about basic hadronic interactions
requires, as a rule, some model, which relates, for example,
proton-nucleus inelastic (absorption) cross section to the
proton-proton total cross section. Having in the hands such reliable
relation we could to discriminate which of the different models for
the high-energy  behaviour of the proton-proton total cross sections
are consistent with cosmic ray data at ultrahigh energies and which
may be ruled out. In this respect, cosmic-ray experiment may serve as
a discriminator or as a filter for the different phenomenological
models.

The procedure generally used to relate the proton-air inelastic cross
section to the proton-proton total cross section is the Glauber
multiple-scattering approach \cite{7}. However, at present time we
can find in the literature some debate concerning the procedure of
extracting the $pp$ total cross sections from cosmic ray experimental
data. We shortly present here the basic conclusions of these
polemics.

It was pointed out in paper \cite{20} (hereafter referred to as NNN)
that Glauber method establishes the relationship between
proton-nucleus absorption cross section and proton-proton inelastic
cross section, which we write in the form
\be
\sigma_{abs}^{p-air}=G[\sigma_{inel}^{pp},B_{el}^{pp},\rho^{nucl}],
\label{1}
\ee
where $\rho^{nucl}$ is nuclear matter density, $B_{el}^{pp}$ is the
slope of $pp$ differential elastic scattering cross section
\be
B_{el}^{pp}= [\frac{d}{dt}(\ln \frac{d\sigma_{el}^{pp}}{dt})]_{t=0},
\label{2}
\ee
$\sigma_{inel}^{pp}$ is $pp$ inelastic cross section
\be
\sigma_{inel}^{pp} = \sigma_{tot}^{pp} - \sigma_{el}^{pp},\label{3}
\ee
$\sigma_{abs}^{p-air}$ is $p-air$ absorption cross section
\be
\sigma_{abs}^{p-air} = \sigma_{tot}^{p-air} - \sigma_{el}^{p-air} -
\sigma_{q-el}^{p-air},\label{4}
\ee
$\sigma_{q-el}^{p-air}$ is quasi-elastic $p-air$ cross section
corresponding to the intermediate excited states of air nucleus, 
$G$ is some known functional of the quantities $\sigma_{inel}^{pp}$,
$B_{el}^{pp}$ and $\rho^{nucl}$.

The same Glauber transformation between inelastic proton-nucleus
cross section and proton-proton total cross section is true as well
\cite{20}
\be
\sigma_{inel}^{p-air}=G[\sigma_{tot}^{pp},B_{el}^{pp},\rho^{nucl}],
\label{5}
\ee
where $G$ is the same functional as in Eq. 1, and
\be
\sigma_{inel}^{p-air} = \sigma_{tot}^{p-air} - \sigma_{el}^{p-air}.
\label{6}
\ee
If we additionally suppose the geometric scaling\footnote{A more
detail discussion an experimental status of geometrical scaling see
in Ref. \cite{22}.} in the form
$\sigma_{inel}^{pp} \sim  B_{el}^{pp}$, then Glauber formula
(\ref{1}) can be used for extraction $\sigma_{inel}^{pp}$ from
measured values of $\sigma_{abs}^{p-air}$. In that case the total
cross section $\sigma_{tot}^{pp}$ is obtained from that extracted
inelastic $pp$ cross section $\sigma_{inel}^{pp}$ by adding the
elastic cross section: $\sigma_{tot}^{pp} = \sigma_{inel}^{pp} +
\sigma_{el}^{pp}$. That is why, Nikolaev argued that the
underestimated values of $\sigma_{tot}^{pp}$ were inferred in Akeno
Collaboration paper \cite{19}. A reanalysis of the Akeno data made in
NNN paper gives a proton-proton total cross section about $30\,mb$
larger than found in \cite{19}, its the main conclusion presented in
NNN paper \cite{20}.

A quite opposite conclusion we found in paper \cite{21} (hereafter
referred to as BHS). Block et al. faced with the problem to predict
proton-air and proton-proton cross sections at energies near
$\sqrt{s}=30\,TeV$ using Glauber approach and their QCD-ispired
parameterization of all accelerator data on forward proton-proton and
proton-antiproton scattering amplitudes. When BHS confronted their
predictions of $p-air$ cross sections $\sigma_{inel}^{p-air}$ as a
function of energy with published cross section measurements of the
Fly's Eye and AGASA groups, they found that the predictions
systematically are about one standard deviation below the published
cosmic ray values \cite{18,19}; see Fig. 1.

\begin{figure}[htb]
\vspace{110pt}
\begin{picture}(120,40)
\put(10,-28){\includegraphics[scale=0.45]{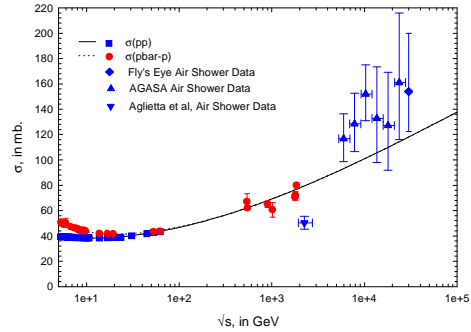}}
\end{picture}
\vspace{-10pt}
\caption[]{\protect
{A plot of the QCD-inspired fit of the nucleon-nucleon total cross
section extracted from paper \cite{21}.}}
\label{fig:1}
\end{figure}
To overcome these troubles Block et al. emphasized that the measured
quantity in cosmic ray experiment is the shower attenuation length or
the mean free path for development of air showers $\Lambda$ which is
not only sensitive to the interaction length of the protons in the
atmosphere (mean free path) $\lambda_{p-air}$ but also depends on the
inelasticity parameter $k$, which determines the rate at which the
energy of the primary proton is dissipated into electromagnetic
shower energy observed in the experiment \cite{21}
\be
\Lambda = k\lambda_{p-air} = k\frac{14.5m_p}{\sigma_{inel}^{p-air}}.
\label{7}
\ee
The rate of shower development and its fluctuations are the origin of
the deviation of $k$ from unity in Eq. (\ref{7}). The value of $k$ is
model dependently estimated through Monte Carlo simulations, its
predicted values range from 1.5 for a model where the inclusive cross
section exhibits Feynman scaling to 1.1 for the models with large
scaling violations. Akeno Collaboration used $k=1.5$ and this value
of $k$ was obtained with the assumption that there is no significant
break of Feynman scaling in the fragmentation region ($x\geq 0.05$)
and that the multiplicity increases as $ln^{2}s$ \cite{19}. If we
assume a breakdown of scaling in the fragmentation region, a smaller
value of $k$ is expected. Fly's Eye Collaboration \cite{18} used
$k=1.6$ with uncertainty of 10\%. 

The extraction of the $pp$ cross section from the cosmic ray data is
a two stage procedure. First, from measured value of $\Lambda$ and
fixed value of $k$ one calculates the $p-air$ inelastic cross section
inferred in Eq. (\ref{7}), where \footnote{It should be born in mind
that the different notations for one and the same quantity have been
used in NNN and BHS papers:
$\sigma_{abs}^{p-air}(NNN)\equiv\sigma_{inel}^{p-air}(BHS)$.}
\[
\sigma_{inel}^{p-air} = \sigma_{tot}^{p-air} - \sigma_{el}^{p-air} -
\sigma_{q-el}^{p-air}. 
\]
This step neglects the possibility that $k$ may have a weak energy
dependence over the range measured.

In the next step the Glauber formula (\ref{1}) transforms the value
of $\sigma_{inel}^{p-air}$ into a proton-proton inelastic cross
section $\sigma_{inel}^{pp}$. Here all the necessary steps are
calculable in the framework of Glauber theory, but depend sensitively
on a knowledge of the slope $B_{el}^{pp}$ as it was mentioned above.

Block et al. decided to let $k$ be a free parameter and to make a
global fit to the accelerator and cosmic ray data using the
QCD-inspired parameterization of the forward proton-proton and
proton-antiproton scattering amplitudes. So, in they global fit, all
4 quantities, $\sigma_{tot}^{pp}$, $B_{el}^{pp}$, $\rho = Re/Im$ and
$k$ were simultaneously fitted. The fit also neglected the energy
dependence of $k$. It was found that the accelerator and cosmic ray
$pp$ cross sections are readily reconcilable using a value of $k =
1.349 \pm 0.045 \pm 0.028$, where the quoted errors are statistical
and systematic ones respectively. They concluded that this
determination of $k$ severely constrains any model of high energy
hadronic interactions.

At the LHC ($\sqrt{s}=14\,TeV$), they predicted $\sigma_{tot}^{pp} =
107.9 \pm 1.2\,mb$ for the $pp$ total cross section, $B_{el}^{pp} =
19.59 \pm 0.11\,(GeV/c)^{-2}$ for the elastic slope and $\rho = 0.117
\pm 0.001$ for the ratio $Re/Im$, where the quoted errors are due to
the statistical errors of the fitting parameters.

\section{Cosmic-ray experiments and theory}

Recently a simple theoretical formula describing the global structure
of $pp$ and $p\bar p$ total cross sections in the whole range of
energies available at now working accelerators has been derived
\cite{5,17}. The fit to the accelerators experimental data with the
formula has been made and it was shown that there is a very good
correspondence of the theoretical formula to the existing
experimental data. In Figs. 2,3, we have presented the fit results.

\begin{figure}[htb]
\vspace{20pt}
\begin{picture}(150,82)
\put(15,-5){\includegraphics[scale=0.65]{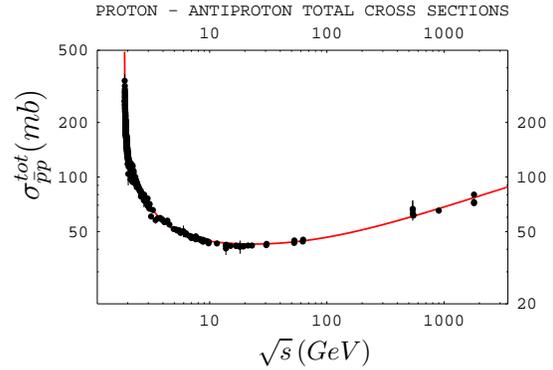}}
\put(92,-15){$\sqrt{s}\, (GeV)$}
\put(0,45){\rotatebox{90}{\large$\sigma^{tot}_{\bar{p}p} (mb)$}}
\end{picture}
\caption[]{\protect
{The proton-antiproton total cross sections
versus $\sqrt{s}$ compared with the theory. Solid line represents our
fit to the data \cite{5,17}. Statistical and systematic errors added
in quadrature.}}
\label{fig:2}
\end{figure}

\begin{figure}[htb]
\vspace{20pt}
\begin{picture}(150,82)
\put(15,-5){\includegraphics[scale=0.65]{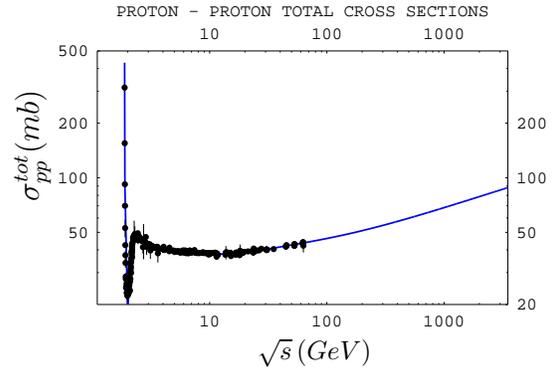}}
\put(92,-15){$\sqrt{s}\, (GeV)$}
\put(0,45){\rotatebox{90}{\large$\sigma^{tot}_{pp} (mb)$}}
\end{picture}
\caption[]{\protect
{The proton-proton total cross sections versus
$\sqrt{s}$ compared with the theory. Solid line represents our fit
to the data \cite{5,17}. Statistical and systematic errors added in
quadrature.}}
\label{fig:3}
\vspace{-15pt}
\end{figure}

It was also demonstrated in papers \cite{5,17} (see Figs. 8,7 there),
that experimental point from cosmic ray Fly's Eye Collaboration do
not contradict to the theoretical predictions made from the fit to
the accelerator data only. Unfortunately, we did not have in the
hands at that time the numerical experimental values from cosmic ray
experiment of AGASA Collaboration. Now these values are available in
the database of Particle Data Group \cite{23} and we can compare our
theoretical predictions with all existing cosmic-ray data on
proton-proton total cross sections. The comparison is shown in Figs.
4,5.

\begin{figure}[htb]
\vspace{20pt}
\begin{picture}(150,82)
\put(15,-5){\includegraphics[scale=0.65]{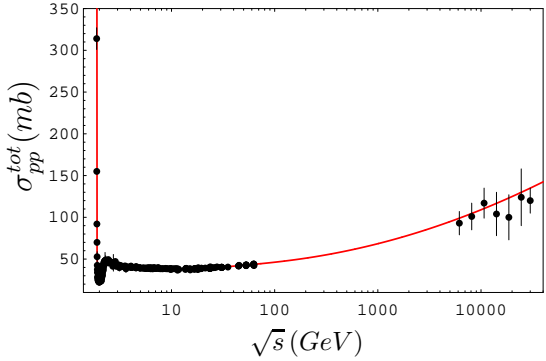}}
\put(92,-15){$\sqrt{s}\, (GeV)$}
\put(0,45){\rotatebox{90}{\large$\sigma^{tot}_{pp} (mb)$}}
\end{picture}
\caption[]{\protect
{The proton-proton total cross-section versus
$\sqrt{s}$ with the cosmic-ray data points from Akeno Observatory
and Fly's Eye Collaboration. Solid line corresponds to our theory
predictions \cite{5,17}.}}
\label{fig:4}
\end{figure}

\begin{figure}[htb]
\vspace{20pt}
\begin{picture}(150,82)
\put(15,-5){\includegraphics[scale=0.65]{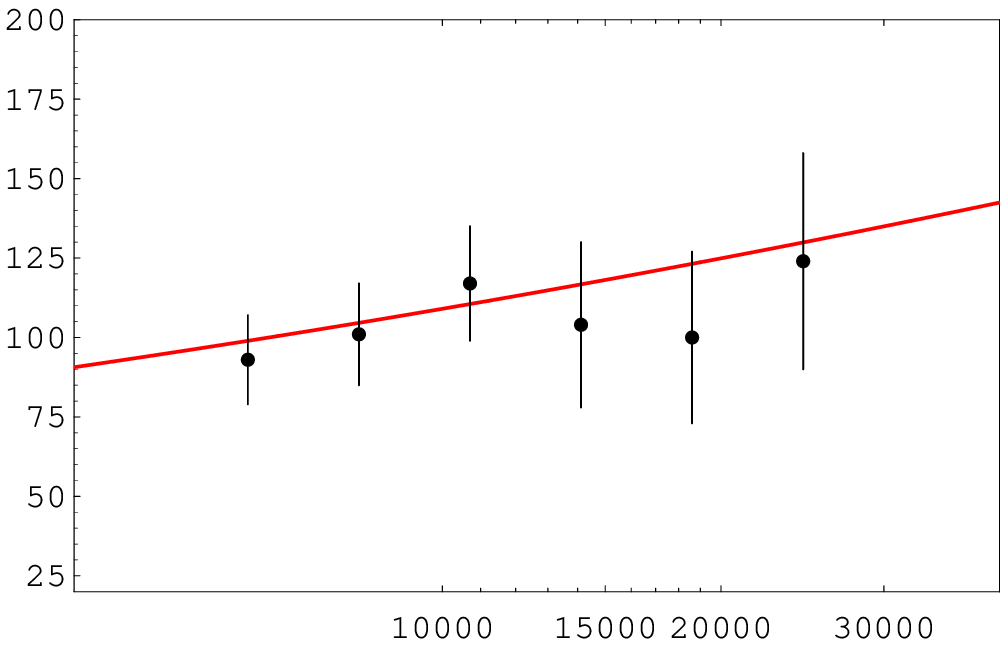}}
\put(92,-15){$\sqrt{s}\, (GeV)$}
\put(0,45){\rotatebox{90}{\large$\sigma^{tot}_{pp} (mb)$}}
\end{picture}
\caption[]{\protect
{The proton-proton total cross-section versus
$\sqrt{s}$ with the cosmic-ray data points from Akeno Observatory
only. Solid line corresponds to our theory predictions \cite{5,17}.}}
\label{fig:5}
\vspace{-15pt}
\end{figure}
As is seen from the Figures
there is very good correspondence of the theory to all existing
cosmic ray experimental data without any reanalysis of the data. What
can we learn from this very nice, at least for us, fact and what
really could it mean?

To understand it more clearly we plotted in Fig. 6 an error band
where upper and lower curves correspond to one deviation in the
fitting parameter $a_2$ which controls the high energy asymptotic in
the total cross section \cite{17}. 

\begin{figure}[htb]
\vspace{20pt}
\begin{picture}(150,82)
\put(15,-5){\includegraphics[scale=0.65]{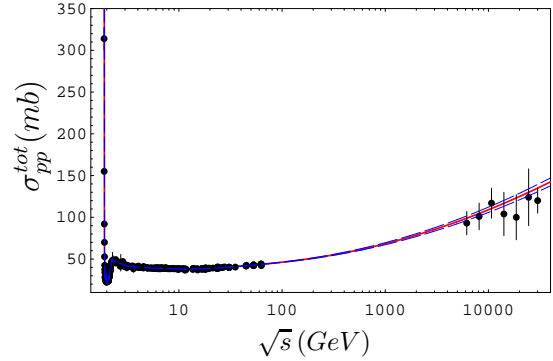}}
\put(92,-15){$\sqrt{s}\, (GeV)$}
\put(0,45){\rotatebox{90}{\large$\sigma^{tot}_{pp} (mb)$}}
\end{picture}
\caption[]{\protect
{The proton-proton total cross-section versus
$\sqrt{s}$ with the cosmic-ray data points from Akeno Observatory
and Fly's Eye Collaboration. Solid line corresponds to our theory
predictions \cite{5,17}. Upper and lower dashed lines show error band
corresponding to one deviation in fitting parameter $a_2$ which
controls the high-energy asymptotic in the total cross section.}}
\label{fig:6}
\vspace{-15pt}
\end{figure}
As one can see from this Figure
the error band is narrow enough so, there is no a large room for the
experimental uncertainties. In this respect a more precise total
cross section $\sigma_{tot}^{pp}$ measurements at cosmic ray energies
are very desirable. Anyway, we would like to emphasize that we faced
here a happy case when the predicted values for  $\sigma_{tot}^{pp}$
obtained from theoretical description of all existing accelerators
data are completely compatible with the values obtained from cosmic
ray experiments. Here we confront with the conclusion made in paper
\cite{24}. 
The best fit of accelerators data made in paper \cite{16}
and they predictions up to cosmic-ray energies are close to our
theory predictions within error band but a little bit lower though.
At the LHC we predict
\be
\sigma_{tot}^{pp}(\sqrt{s}=14\,TeV) = 116.53 \pm 3.52\,mb,\label{8}
\ee
which is in $3\sigma$ higher than the BHS prediction. Our estimated
value $\sigma_{tot}^{pp}(\sqrt{s}=40\,TeV) = 142.46\,mb$ is
significantly lower than the value predicted by NNN; see Fig. 7.

\begin{figure}[htb]
\vspace{20pt}
\begin{picture}(150,82)
\put(15,-5){\includegraphics[scale=0.65]{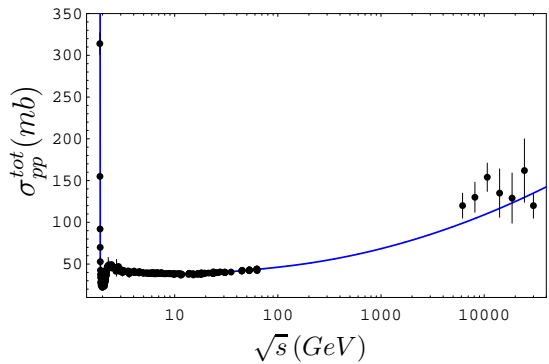}}
\put(92,-15){$\sqrt{s}\, (GeV)$}
\put(0,45){\rotatebox{90}{\large$\sigma^{tot}_{pp} (mb)$}}
\end{picture}
\caption[]{\protect
{The proton-proton total cross-section versus
$\sqrt{s}$ with the cosmic-ray data points  extracted from NNN paper
\cite{20}. Solid line corresponds to our theory predictions
\cite{5,17}.}}
\label{fig:7}
\vspace{-15pt}
\end{figure}

Let's try to explain these discrepancies. In reanalysis of cosmic ray
experimental data made by NNN the formula
\be
\sigma_{in}^{pp} = [\sigma_{abs}^{p-air}/507\,mb]^{1.89}\cdot 100\,mb
\label{9}
\ee
has been used. It has been argued by NNN that this formula is valid
to a few percent accuracy at $\sigma_{abs}^{p-air}>300\,mb$ and/or
$\sigma_{in}^{pp}>37\,mb$. Using cosmic ray experimental data
identified with $\sigma_{abs}^{p-air}$ Nikolaev obtained
$\sigma_{in}^{pp}$ with the help of formula (\ref{9}) and after that
the total cross section $\sigma_{tot}^{pp}$ was obtained by adding
the elastic cross section: 
$\sigma_{tot}^{pp} = \sigma_{in}^{pp} + \sigma_{el}^{pp}$. However,
it should be pointed out that Eq. \ref{9} is not a theoretically well
grounded formula but a pure phenomenological one. That is why, the
further theoretical study of multiparticle dynamics is needed.

Concerning the BHS analysis we might apply the arguments of NNN and
say that, in fact, BHS extracted $\sigma_{in}^{pp}$ from cosmic ray
experimental measurements of $\sigma_{inel}^{p-air}$ but not
$\sigma_{tot}^{pp}$ as it has been argued by BHS. Therefore to obtain
$\sigma_{tot}^{pp}$ we have to add $\sigma_{el}^{pp}$ to the values
extracted by BHS, and in that case we would come to the agreement
with the values published by cosmic-ray experimental groups. It
should be noted that BHS did not cite the paper of NNN.

Our theory predictions and cosmic-ray experimental data are just in
the middle between NNN and BHS. We suppose that this is the Golden
Middle. 

\vspace*{-2mm}
\section{Conclusion}

In conclusion, we would like to hope that in the near future it would
be possible to repeat the cosmic ray experiments to measure (or to
extract) the proton-proton total cross sections with a higher
accuracy. A more precise total cross section $\sigma_{tot}^{pp}$
measurements at cosmic ray energies are very desirable. Surely, we
have in analysing the experimental data to use a right theory, which
in our opinion the local Quantum Field Theory is. Here we did not
make any attempts to rescale the cosmic-ray experimental data to fit
them to our theory predictions or to some phenomenological formula as
it was made in BHS paper. In contrary to BHS we have shown that our
theoreticaly derived formula, which describes the global structure of
all accelerator data for proton-proton total cross sections from the
most low energies to the most high ones, is completely compatible
with the existing  cosmic-ray experimental data as well. This is our
basic conclusion. 
  
\vspace*{-2mm}  
\section*{Acknowledgements}

The material of this paper was presented at the IX Blois Workshop
held in Pruhonice near Prague 9-15 June, 2001. So, I'd like to use
the opportunity to thank the Organizing Committee for the kind
invitation to attend this Workshop. I am also grateful to V.V. Ezhela
(COMPAS Group, IHEP, Protvino) for his courtesy to present the
numerical values of all existing cosmic-ray experimental data on
proton-proton total cross section.

\end{document}